\documentclass[pra,reprint]{revtex4-1}

\usepackage{amsmath}
\usepackage{graphicx}
\usepackage{epsfig}

\def\eref#1{Eq~\ref{#1}}

\newcommand{\vecx}{\ensuremath{\mathbf{x}}}
\newcommand{\extpot}{\ensuremath{V(\vecx)}}

\newcommand{\scatpot}{\ensuremath{V_{\mathrm{S}}}}
\newcommand{\trappotx}{\ensuremath{V_{\mathrm{T}}(\vecx)}}
\newcommand{\scatpotx}{\ensuremath{V_{\mathrm{S}}(\vecx)}}
\newcommand{\density}{\ensuremath{\rho_\mathrm{a}(\vecx)}}
\newcommand{\HH}{\ensuremath{\mathrm{H}}}

\begin{document}
\date{\today}
\author{K. J. Arnold, M. P. Baden, M. D. Barrett}
\affiliation{ Centre for Quantum Technologies and Department of
  Physics, National University of Singapore, 3 Science Drive 2, 117543 Singapore}
\title{Self-Organization Threshold with an External Potential}
\begin{abstract}
We derive threshold equations for self-organization of laser driven atoms in an optical cavity. Our analysis includes probing with either a traveling wave or a retro reflected lattice. These two scenarios lead to qualitatively different behavior in terms of the response of the system as a function of cavity detuning with respect to the probe.  In addition our analysis includes the effects of an intra-cavity trapping potential which is also shown to impact on the threshold condition.  We specifically consider the case of an intra-cavity lattice as used in \cite{Kyle2} but our treatment can easily be modified to other geometries.
\end{abstract}

\maketitle
\section{Introduction}
Atoms coupled to the standing wave mode of a cavity will self-organize when driven by a sufficiently intense transverse laser field \cite{ritsch2002}. As the atoms scatter more light into the cavity, a potential is produced which localizes the atoms in a more favorable configuration for scattering. This further enhances the localization of the atoms. Thus above a threshold pump intensity, an initially uniform distribution of atoms will undergo a phase transition, spontaneously reorganizing into a lattice configuration.  Equations determining the threshold intensity are given in \cite{vukics1} but they are specific to a set up in which the atoms are driven by a standing wave field.  They are also restricted to a particular probe detuning, and an initial distribution of atoms in which its transverse extent can be neglected.  Here we extend the analysis to include the effects of an external confinement of the atoms and a traveling wave probing configuration.  We restrict our attention to a lattice confinement as used in recent experiments \cite{Kyle2} but the analysis can easily be modified to other trapping configurations.  For the case of a traveling wave probe it is found that the threshold behavior is qualitatively different to that of a retroreflected lattice geometry and is analogous to a collective atom recoil laser (CARL) \cite{kruse2003}.  We first give an overview of the self-organization process introducing the needed terms and equations for our analysis.  We then give a detailed derivation of the threshold conditions for both the traveling wave and retro-reflected probing conditions.

\section{Self-organization of thermal atoms in external potentials}

To derive the equations for the onset of self-organization of atoms coupled to an optical cavity, we extend the previous theoretical treatment of \cite{vukics1} to include an external trapping potential and the spatial extent of the atomic distribution transverse to the cavity. In thermal equilibrium the spatial density distribution of atoms in a confining potential $\extpot$ is given by
\begin{equation}
\label{Boltzmann}
\density =\frac{1}{Z}\exp\left(-\frac{\extpot}{k_B T}\right),
\end{equation}
where $Z=\int\exp\left(-\extpot)/(k_B T)\right)\,\mathrm{d}^3 x$. For our case the confining potential consists of two components
\begin{equation}
  \label{eq:1}
  \extpot = \trappotx + \scatpotx,
\end{equation}
where $\trappotx$ is the potential associated with an external potential and $\scatpotx$ is the potential associated with the probe beam itself and light scattered from the probe beam into the cavity.

Self-organization in this system arises from the back-action of the scattered light on the atomic distribution.  As the atoms scatter light into the cavity, the resulting potential $\scatpotx$ localizes the atoms to a more favorable configuration for scattering. This results in increased scattering which then further enhances the localization of the atoms, and, under sufficient driving strength, an initially uniform distribution of atoms will undergo a phase transition spontaneously reorganizing into one of two possible lattice configurations.

In our analysis of the problem two spatial averages play an important role, namely
\begin{subequations}
\label{averages}
\begin{align}
\label{dshift}
\alpha &= \int f^2(r,z) \rho_a(\mathbf{x}) \,\mathrm{d}^3x\\
\intertext{and}
\label{order}
\beta &= \int f(r,z) h(x) \rho_a(\mathbf{x}) \,\mathrm{d}^3x,
\end{align}
\end{subequations}
where, $f(r,z)=\cos(k z) \exp(-r^2/w^2)$ is the mode function of the cavity and $h(x)$ the mode function of the probe beam, which in our experiments has either a standing or traveling wave form. The factor $\alpha$ accounts for the reduced coupling of the ensemble to the cavity due the spatial positions of the individual atoms relative to the cavity mode. The factor $\beta$ quantifies the scattering into cavity due to the spatial organization of the atoms. It takes the role of an order parameter having a value near zero prior to the phase transition and rapidly approaches $\pm 1$ as the probe coupling, $\Omega$, rises above a threshold value.

Since the potential $\scatpot$ in general is a function of $\alpha$ and $\beta$, \eref{Boltzmann} and \ref{averages} must be solved self-consistently.  This can be done by first expressing $\scatpot$ in terms of arbitrary $\alpha$ and $\beta$ and using \eref{Boltzmann} to determines the atomic distribution.  The distribution function is then substituted into \eref{averages} to give a coupled set of self-consistency equations for $\alpha$ and $\beta$.  To determine the threshold condition for self-organization to occur we note that, near the critical point, $\beta\approx 0$. Hence we can expand the self-consistency equations to lowest non-zero order in $\beta$.  This yields the necessary condition for there to be a non-zero value of $\beta$.

We first determine $\scatpotx$ by looking at the steady
state solutions describing the system for arbitrary $\alpha$ and
$\beta$. At large atomic detunings, we can adiabatically
eliminate the atomic excited states. The resulting Hamiltonian
for a collection of two-level atoms with a fixed density, $\density$,
inside a cavity driven from the side by a classical field is
\begin{equation}
  \label{hamiltonian}
  \HH = (-\Delta_c + N U_0 \alpha) a^\dagger a+N (\Omega_R \beta)^* a+N (\Omega_R \beta) a^\dagger
\end{equation}
with the cavity decay described by the Liouvillian
\begin{equation}
\mathcal{L}\rho=(\kappa+N \Gamma_0 \alpha)(2 a\rho a^\dagger-a^\dagger a \rho-\rho a^\dagger a).
\end{equation}
In these expressions $\Delta_c=\omega_p-\omega_c$ is the cavity detuning with respect to the probe, $U_0$ is the single atom dispersive shift, $\Omega_R$ is the pump rate per atom of the cavity due to scattering of the probe field, and $\Gamma_0$ is a correction to the cavity decay associated with spontaneous emission. In terms of the atom-cavity coupling, $g$, atom-probe coupling, $\Omega$, the detuning, $\Delta$, of the probe from the atomic resonance, and the atomic linewidth, $\gamma$, the parameters $U_0$, $\Gamma_0$, and $\Omega_R$ may be written
\[
U_0 = \frac{g^2}{\Delta}, \quad \Gamma_0 = \gamma \frac{g^2}{\Delta^2}, \quad \Omega_R = \frac{\Omega g}{\Delta}
\]
In our system $N\Gamma_0\ll \kappa$ and so we neglect this term in all that follows.

In steady state the cavity mode is described by a coherent state with amplitude
\begin{equation}
  \label{eq:3}
  \lambda = \frac{N \Omega_R \beta}{\tilde{\Delta}_C + i \kappa},
\end{equation}
where $\tilde{\Delta}_c=\Delta_c-N U_0 \alpha$ is the detuning from the dispersively shifted cavity.

The total scattered potential, $\scatpotx$, is then given by
\begin{align}
 \scatpotx &= \frac{\hbar}{\Delta} |\Omega h(x) + \lambda g f(x) |^2 \nonumber \\
  \begin{split}
    &= \frac{\hbar\Omega^2}{\Delta}\bigg\{|h(x)|^2 + 2 \epsilon f(r,z) \mathrm{Re}\left(\beta^* e^{-i\theta} h(x)\right)\\
     &\quad + \epsilon^2 |\beta|^2 f^2(r,z)\bigg\}
  \end{split}    \label{scattered-potential}
\end{align}
where
\begin{equation}
\label{para1}
\epsilon = -\frac{N U_0}{\sqrt{\tilde{\Delta}_c^{2}+\kappa^2}} \quad \text{and} \quad e^{i \theta}=\frac{-\tilde{\Delta}_c+i \kappa}{\sqrt{\tilde{\Delta}_c^{2}+\kappa^2}}.
\end{equation}
Using the resulting potential $\extpot = \trappotx + \scatpotx$ to determine the density distribution via Eq.~\ref{Boltzmann} will lead to self-consistency equations for $\alpha$ and $\beta$ via their definitions in Eq.~\ref{averages}.

\section{Derivation of the threshold equations}
\label{sec:old-thresh-deriv}

The self-consistency equations for $\alpha$ and $\beta$ are given by
\begin{subequations}
\label{averages2}
\begin{equation}
\label{alphaeq}
\alpha=\frac{1}{Z}\int e^{-(\trappotx + \scatpotx)/(k_B T)}f^2(r,z)\,\mathrm{d}^3 x
\end{equation}
and
\begin{equation}
\label{betaeq}
\beta=\frac{1}{Z}\int e^{-(\trappotx + \scatpotx)/(k_B T)} f(r,z) h(x)\,\mathrm{d}^3 x
\end{equation}
\end{subequations}
with
\begin{equation}
Z=\int e^{-(\trappotx + \scatpotx)/(k_B T)}\,\mathrm{d}^3 x.
\end{equation}
We henceforth consider the specific case in which $V_T$ is given by an intra-cavity FORT that has a wavelength twice that of the probe as used in recent experiments \cite{Kyle2}.  This traps the atoms at every alternate antinode of the cavity .  In this case, the integration along the $z$ direction can be carried out by integrating over single trapping sites FORT lattice potential and weighting each site in accordance with the probability that the site is occupied.  Since the integrands have the same periodicity as the lattice potential, the integrations over each trapping site are equal and we need only consider a single site.  This is equivalent to considering all the atoms to be located in one site.  Furthermore, we assume the atoms to be sufficiently well localized by the FORT potential so that the trapping site can be well represented by its harmonic approximation.  We thus replace the potential $V_T$ by its harmonic approximation
\[
V_T=\frac{1}{2} m \omega_r^2 r^2+\frac{1}{2} m \omega_z^2 z^2
\]
and extend the $z$ integration out to infinity.

\subsection{Lattice geometry}
In the case of probing the system with a lattice, the mode
function is
\begin{equation}
  \label{eq:lattice-mode}
  h(x)=\cos(kx-\phi)
\end{equation}
where $k = 2 \pi / \lambda $ and $\lambda$ the wavelength  of the probe light.  The potential $V_s$ is then given by
\begin{eqnarray}
\label{lattice}
V_s &=& \frac{\hbar\Omega^2}{\Delta}\bigg[\cos^2(k x-\phi)\nonumber\\
& & \hspace{1.5cm}+ 2 \epsilon \beta f(r,z) \cos(k x-\phi) \cos(\theta)\bigg]
\end{eqnarray}
where $\epsilon$ and $\theta$ are determined by \eref{para1}. Note
that we have dropped a term proportional to $\beta^2$ as we will only
be interested in terms up to first order. When we expand \eref{averages2} to lowest non-zero order in $\beta$ we encounter integrals of the form
\[
\mathcal{I}=\int_{-\infty}^\infty e^{-x^2/(2\sigma^2)}W(\cos(kx-\phi))\,\mathrm{d} x
\]
for some function $W$. Here $\sigma$ is approximately the spatial extent of the atomic distribution along $x$. These integrals can be greatly simplified when $k\sigma \gg 1$ which means the spatial extent of the atomic distribution is larger than the wavelength.  In this case the integrand can be averaged over the wavelength. Denoting the averaging by $\langle \rangle$ we then have
\begin{eqnarray}
\mathcal{I} & \approx &\left\langle W(\cos(kx-\phi))\right\rangle \int_{-\infty}^\infty e^{-x^2/(2\sigma^2)}\,\mathrm{d} x\nonumber\\
&=& \left\langle W(\cos(kx))\right\rangle \sigma \sqrt{2\pi}
\end{eqnarray}
Thus, within this approximation, the phase $\phi$ of the lattice potential is irrelevant and we subsequently set it to zero.  The expansions of \eref{averages2} can now be readily evaluated.

The zeroth order expansion of \eref{alphaeq} gives
\begin{eqnarray}
\alpha &=& \frac{\int e^{-V_T/(k_B T)}e^{\mu \cos^2(kx)}f^2(r,z)\,\mathrm{d}^3 x}{\int e^{-V_T/(k_B T)}e^{\mu \cos^2(kx)}\,\mathrm{d}^3 x}\nonumber\\
&=& \frac{\int e^{-V_T/(k_B T)}f^2(r,z)\,\mathrm{d}^3 x}{\int e^{-V_T/(k_B T)}\,\mathrm{d}^3x}
\end{eqnarray}
where
\begin{equation*}
\mu = -\hbar \Omega^2 / (\Delta k_B T)
\end{equation*}
is the depth of the potential due to the probe relative to the thermal energy $k_B T$ of the atoms. Within the harmonic approximation for $\trappotx$ we then find
\begin{equation}
\label{alpha}
\alpha=\frac{1}{2}\frac{1+e^{-4/\eta}}{1+2/\eta}.
\end{equation}
where $\eta=V_{T0}/(k_B T)$ and $V_{T0}$ is the depth of the FORT potential.  Note we have utilized the fact that the waist of the FORT beam is $\sqrt{2}$ larger than that of the cavity mode associated with the probe wavelength.

For the expansion of \eref{betaeq}, the zeroth order term is proportional to an integral of the form
\begin{equation*}
\int_{-\infty}^\infty e^{-x^2/(2\sigma^2)}e^{\mu \cos^2(kx)}\cos(kx)\,\mathrm{d}x.
\end{equation*}
Within the approximation $k\sigma\gg 1$ this term is zero as expected for nearly uniform distribution per wavelength.  If the approximation $k\sigma\gg 1$ is not satisfied then the zeroth order term would be non-zero and $\beta$ would be non-zero for any value of the probe coupling.  This is because the atoms would already have sufficient localization to provide some level of scattering into the cavity for any probe intensity and thus no threshold would exist.

The expansion of \eref{betaeq} to first order then gives
\begin{eqnarray}
\beta &=& 2\epsilon \mu \beta \cos(\theta) \nonumber\\
& & \times \frac{\int e^{-V_T/(k_B T)}e^{\mu \cos^2(kx)}\cos^2(kx)f^2(r,z)\,\mathrm{d}^3 x}{\int e^{-V_T/(k_B T)}e^{\mu \cos^2(kx)}\,\mathrm{d}^3 x}\nonumber\\
&=& 2\epsilon \mu \beta \cos(\theta) \alpha \frac{\left\langle e^{\mu \cos^2(kx)}\cos^2(kx)\right\rangle}{\left\langle e^{\mu \cos^2(kx)}\right\rangle}\nonumber\\
&=& \left[\epsilon \mu \cos(\theta)\alpha \left(1+\frac{I_1(\mu/2)}{I_0(\mu/2)}\right)\right] \beta
\end{eqnarray}
A nontrivial solution for $\beta$ requires the expression within the square brackets to be unity thus giving the required threshold equation
\begin{equation}
\label{threshold1}
\left(1+\frac{I_1(\mu/2)}{I_0(\mu/2)}\right)\mu =\frac{1}{N U_0 \alpha}\frac{\tilde{\Delta}_c^2+\kappa^2}{\tilde{\Delta}_c}.
\end{equation}

It is worthwhile comparing our result to the original treatment in \cite{vukics1}. Our threshold equation is derived under the assumption that the atoms are trapped at every second antinode of the cavity and have a transverse spatial extent that is large relative to the wavelength.  The result given in ~\cite{vukics1} on the other hand neglects the transverse dimension and assumes a uniform distribution along the z-axis.  However these differences do not significantly impact on the threshold equation.  Confining the atoms to every second antinode merely alters the parameter $\alpha$ and the net effect is a slight rescaling of the threshold value.  Including the spatial distribution of the atoms transverse to the cavity gives rise to the ratio $I_1(\mu/2)/I_0(\mu/2)$ in the threshold equation.  This ratio should be set to $\approx 1$ if the atoms are confined to length scales much smaller than the wavelength along this dimension. Thus recovering Eq (19) of ref~\cite{vukics1} amounts to setting $\alpha=1/2$ to account for the initially uniform distribution of atoms along the cavity axis, and setting $I_1(\mu/2)/I_0(\mu/2)=1$ to account for neglecting the transverse dimension.  Using their chosen detuning of $\tilde{\Delta}_c=-\kappa$ then gives exactly the expression reported in ref~\cite{vukics1}.  Our threshold equation is therefore a generalization of the result given in \cite{vukics1} to account for an arbitrary detuning of the cavity and for the transverse extent of the atomic distribution.

\subsection{Traveling wave geometry}
When probing with a travelling wave beam, the mode function is
\begin{equation}
  \label{eq:4}
  h(x)=e^{-i k x}
\end{equation}
and the potential $V_s$ is given by
\begin{eqnarray}
\label{TW}
V_s &=& \frac{\hbar\Omega^2}{\Delta}\left[2 \epsilon f(r,z) \mathrm{Re}\left(\beta^*e^{i(kx-\theta)}\right)\right]\nonumber\\
&=& \frac{\hbar\Omega^2}{\Delta}\left[2 \epsilon f(r,z) |\beta|\cos(kx-\bar{\theta})\right]
\end{eqnarray}
where $\bar{\theta}=\theta+\arg(\beta)$.  Note that we have dropped the constant term $\hbar\Omega^2/\Delta$ and as before we have neglected the term proportional to $\beta^2$ as we are only interested in terms up to first order.

Taking the zeroth order expansion of \eref{alphaeq} we get exactly the same expression for $\alpha$ as we do for the lattice case.  This is not surprising as $\alpha$ represents the averaging of the dispersive shift due to the spatial distribution of the atoms.  As we assume the spatial extent in the transverse direction is much larger than the wavelength this averaging will be unaffected even if a lattice potential is invoked along this dimension.

For the same reason as for the lattice case, the zeroth order term in the expansion of \eref{betaeq} vanishes and to first order we have
\begin{eqnarray}
\label{betaeqfinal}
\beta &=& 2\epsilon \mu |\beta|\frac{\int e^{-V_T/(k_B T)} f^2(r,z) e^{i k x} \cos(kx-\bar{\theta}) \,\mathrm{d}^3 x}{\int e^{-V_T/(k_B T)}\,\mathrm{d}^3 x}\nonumber\\
&=& 2\epsilon \mu |\beta|\alpha \left\langle e^{i k x} \cos(kx-\bar{\theta}) \right\rangle\nonumber\\
&=& \epsilon \mu |\beta|\alpha e^{i\bar{\theta}}\nonumber\\
&=& \left(\epsilon \mu \alpha e^{i\theta}\right)\beta.
\end{eqnarray}
Equating the term in parentheses to unity then yields the equation
\begin{equation}
\label{threshold2}
\mu=\frac{\sqrt{\tilde{\Delta}_c^2+\kappa^2}}{-N U_0 \alpha} e^{i\theta}.
\end{equation}
However, since $\mu$ is real, this equation has no solution due to the $e^{i\theta}$ term.  The difficulty stems from the fact that, for an organized array of atoms in the cavity, the scattered potential has minima that are displaced from the positions of the atoms. The result would be that, as the transverse lattice potential forms, the atoms would experience a force along the direction of the probe, establishing a moving lattice similar to the situation encountered in the collective atom recoil laser (CARL).  We therefore argue that the appropriate threshold equation for this geometry is \eref{threshold2} with the $e^{i\theta}$ term omitted.  We also note if an appropriate force were applied to the atoms, they would be held at a position offset from the potential minima giving a self consistent solution for $\alpha$ and $\beta$. Such a force could be provided by the Gaussian profile of the cavity mode and would have the effect of stabilizing the lattice by canceling the phase term in \eref{threshold2}.
\section{Conclusion}
We have derived threshold conditions governing the onset of self-organization for $N$ atoms in a high finesse cavity.  Our equations generalize those given in \cite{vukics1} by allowing for arbitrary detuning of the probe relative to the cavity, and account for the full spatial extent of the initial atomic distribution.  We have also included the case of a traveling wave probe that has not yet been considered.  In this case the response of the system is qualitatively different from that of a lattice probe.  Specifically, self-organization occurs for a probe detuned above the cavity resonance contrary to the lattice case.  Our derivations have been restricted to the specific case which includes an intra-cavity FORT but the treatment can be easily modified to other external trapping potentials.


\begin{thebibliography}{4}%
\makeatletter
\providecommand \@ifxundefined [1]{%
 \@ifx{#1\undefined}
}%
\providecommand \@ifnum [1]{%
 \ifnum #1\expandafter \@firstoftwo
 \else \expandafter \@secondoftwo
 \fi
}%
\providecommand \@ifx [1]{%
 \ifx #1\expandafter \@firstoftwo
 \else \expandafter \@secondoftwo
 \fi
}%
\providecommand \natexlab [1]{#1}%
\providecommand \enquote  [1]{``#1''}%
\providecommand \bibnamefont  [1]{#1}%
\providecommand \bibfnamefont [1]{#1}%
\providecommand \citenamefont [1]{#1}%
\providecommand \href@noop [0]{\@secondoftwo}%
\providecommand \href [0]{\begingroup \@sanitize@url \@href}%
\providecommand \@href[1]{\@@startlink{#1}\@@href}%
\providecommand \@@href[1]{\endgroup#1\@@endlink}%
\providecommand \@sanitize@url [0]{\catcode `\\12\catcode `\$12\catcode
  `\&12\catcode `\#12\catcode `\^12\catcode `\_12\catcode `\%12\relax}%
\providecommand \@@startlink[1]{}%
\providecommand \@@endlink[0]{}%
\providecommand \url  [0]{\begingroup\@sanitize@url \@url }%
\providecommand \@url [1]{\endgroup\@href {#1}{\urlprefix }}%
\providecommand \urlprefix  [0]{URL }%
\providecommand \Eprint [0]{\href }%
\providecommand \doibase [0]{http://dx.doi.org/}%
\providecommand \selectlanguage [0]{\@gobble}%
\providecommand \bibinfo  [0]{\@secondoftwo}%
\providecommand \bibfield  [0]{\@secondoftwo}%
\providecommand \translation [1]{[#1]}%
\providecommand \BibitemOpen [0]{}%
\providecommand \bibitemStop [0]{}%
\providecommand \bibitemNoStop [0]{.\EOS\space}%
\providecommand \EOS [0]{\spacefactor3000\relax}%
\providecommand \BibitemShut  [1]{\csname bibitem#1\endcsname}%
\let\auto@bib@innerbib\@empty
\bibitem [{\citenamefont {Arnold}\ \emph {et~al.}(2012)\citenamefont {Arnold},
  \citenamefont {Baden},\ and\ \citenamefont {Barrett}}]{Kyle2}%
  \BibitemOpen
  \bibfield  {author} {\bibinfo {author} {\bibfnamefont {K.~J.}\ \bibnamefont
  {Arnold}}, \bibinfo {author} {\bibfnamefont {M.~P.}\ \bibnamefont {Baden}}, \
  and\ \bibinfo {author} {\bibfnamefont {M.~D.}\ \bibnamefont {Barrett}},\
  }\href@noop {} {\  (\bibinfo {year} {2012})},\ \Eprint
  {http://arxiv.org/abs/1205.4186v1} {arXiv:1205.4186v1} \BibitemShut {NoStop}%
\bibitem [{\citenamefont {Domokos}\ and\ \citenamefont
  {Ritsch}(2002)}]{ritsch2002}%
  \BibitemOpen
  \bibfield  {author} {\bibinfo {author} {\bibfnamefont {P.}~\bibnamefont
  {Domokos}}\ and\ \bibinfo {author} {\bibfnamefont {H.}~\bibnamefont
  {Ritsch}},\ }\href {\doibase 10.1103/PhysRevLett.89.253003} {\bibfield
  {journal} {\bibinfo  {journal} {Phys. Rev. Lett.}\ }\textbf {\bibinfo
  {volume} {89}},\ \bibinfo {pages} {253003} (\bibinfo {year}
  {2002})}\BibitemShut {NoStop}%
\bibitem [{\citenamefont {Asb\'oth}\ \emph {et~al.}(2005)\citenamefont
  {Asb\'oth}, \citenamefont {Domokos}, \citenamefont {Ritsch},\ and\
  \citenamefont {Vukics}}]{vukics1}%
  \BibitemOpen
  \bibfield  {author} {\bibinfo {author} {\bibfnamefont {J.~K.}\ \bibnamefont
  {Asb\'oth}}, \bibinfo {author} {\bibfnamefont {P.}~\bibnamefont {Domokos}},
  \bibinfo {author} {\bibfnamefont {H.}~\bibnamefont {Ritsch}}, \ and\ \bibinfo
  {author} {\bibfnamefont {A.}~\bibnamefont {Vukics}},\ }\href {\doibase
  10.1103/PhysRevA.72.053417} {\bibfield  {journal} {\bibinfo  {journal} {Phys.
  Rev. A}\ }\textbf {\bibinfo {volume} {72}},\ \bibinfo {pages} {053417}
  (\bibinfo {year} {2005})}\BibitemShut {NoStop}%
\bibitem [{\citenamefont {Kruse}\ \emph {et~al.}(2003)\citenamefont {Kruse},
  \citenamefont {von Cube}, \citenamefont {Zimmermann},\ and\ \citenamefont
  {Courteille}}]{kruse2003}%
  \BibitemOpen
  \bibfield  {author} {\bibinfo {author} {\bibfnamefont {D.}~\bibnamefont
  {Kruse}}, \bibinfo {author} {\bibfnamefont {C.}~\bibnamefont {von Cube}},
  \bibinfo {author} {\bibfnamefont {C.}~\bibnamefont {Zimmermann}}, \ and\
  \bibinfo {author} {\bibfnamefont {P.~W.}\ \bibnamefont {Courteille}},\ }\href
  {\doibase 10.1103/PhysRevLett.91.183601} {\bibfield  {journal} {\bibinfo
  {journal} {Phys. Rev. Lett.}\ }\textbf {\bibinfo {volume} {91}},\ \bibinfo
  {pages} {183601} (\bibinfo {year} {2003})}\BibitemShut {NoStop}%
\end{thebibliography}

%

\end{document}